# H$_2$-Cooling of Primordial Gas Triggered by UV Irradiation

Zoltán Haiman[1,2], Martin J. Rees[2], and Abraham Loeb[1]

## ABSTRACT

We investigate the formation of molecular hydrogen (H$_2$) in a primordial H+He gas cloud irradiated by a power-law UV flux. We find that at high densities ($\gtrsim 1$ cm$^{-3}$) and low temperatures ($\lesssim 10^4$ K), the background radiation enhances the formation of H$_2$ and results in molecular cooling dominating over photoionization heating. This process could accelerate the collapse and fragmentation of dense objects at high redshifts.

*Subject headings:* cosmology:theory-early universe-molecular processes-galaxies:formation

## 1. Introduction

It has long been hypothesized, based on studies of QSO absorption spectra (Gunn & Peterson 1965), that the intergalactic gas is photoionized by an intense UV background. Indeed, a background flux level $\sim 10^{-21\pm 1}$ erg cm$^{-2}$ Hz$^{-1}$ sr$^{-1}$ at the Lyman limit has been inferred from the proximity effect in the vicinity QSOs at redshifts $z \approx 2-4$ (see Bechtold 1994, and references therein). This UV background could have resulted from either QSO or stellar emission (e.g., Miralda-Escudé & Ostriker 1990; Meiksin & Madau 1993; Shapiro, Giroux, & Babul 1994; Giroux, Fardal, & Shull 1995). Aside from its impact on the ionization state of the universe, the UV background heats the intergalactic gas and therefore raises its Jeans mass. Couchman & Rees (1986) realized the important negative feedback that this process might have on the formation of low-mass objects at high redshift. More recently, Efstathiou (1992) suggested that photoionization heating of the intergalactic medium to temperatures $\gtrsim 10^4$ K could have suppressed the formation of dwarf galaxies ($\sim 10^9 M_\odot$) at moderate redshifts, and Babul & Rees (1992) have argued that the formation epoch of faint blue galaxies was indeed delayed by this process until $z \lesssim 1$ (see also Babul & Ferguson 1995). The negative feedback that photoionization heating has on structure formation on small mass scales was verified recently by a number of numerical simulations (Hernquist et al. 1995; Quinn, Katz, & Efstathiou 1995; Steinmetz 1995; Thoul & Weinberg 1995).

---

[1]Astronomy Department, Harvard University, 60 Garden Street, Cambridge, MA 02138, USA

[2]Institute of Astronomy, Cambridge University, Madingley Road, Cambridge CB3 1RZ, England

Naively, one would expect that a relatively cold gas cloud ($\lesssim 10^4$ K) could only heat due to its interaction with the ionizing background. However, this expectation ignores the formation of molecular hydrogen ($H_2$). In this *Letter* we show that the increased abundance of free electrons due to photoionization tends to catalyze the formation of molecular hydrogen, which at sufficiently high densities results in a net *cooling* of the gas. To demonstrate this effect we consider a stationary homogeneous gas cloud which is being illuminated by an ionizing flux. §2 describes our method of calculation and §3 presents our results. Finally, we briefly sketch the potential cosmological implications from this effect in §4.

## 2. Method of Calculation

We consider a homogeneous gas cloud of primordial composition at a density $\rho = \mu m_\mathrm{p} n$ and temperature $T$, made up of the nine species H, $H^-$, $H^+$, He, $He^+$, $He^{++}$, $H_2$, $H_2^+$, and $e^-$. The chemical composition of the gas changes in time according to the rate equations

$$\frac{dn_\mathrm{i}}{dt} = \sum_{l=1}^{9} \sum_{m=1}^{9} a_\mathrm{lmi} k_\mathrm{lm} n_\mathrm{l} n_\mathrm{m} + \sum_{j=1}^{9} b_\mathrm{ji} k_\mathrm{j} n_\mathrm{j}, \qquad (1)$$

where $n_\mathrm{i}$ is the number density of the $i^\mathrm{th}$ species, $k_\mathrm{lm}$ denote reaction rate coefficients, $k_\mathrm{n}$ denote photo-ionization or photo-dissociation integrals, and the factors $a_\mathrm{lmi}, b_\mathrm{ni} = 0, \pm 1$, or $\pm 2$, depending on the reaction. The full list of reactions and rate coefficients that we adopt are given in Appendix B of Haiman, Thoul, & Loeb 1995.

We assume that the gas is illuminated by a uniform background radiation with a power-law flux extending up to 40keV,

$$J_\nu = J_{21} \left(\frac{\nu}{\nu_\mathrm{H}}\right)^{-\alpha} \times 10^{-21} \ \mathrm{erg\ s^{-1}\ cm^{-2}\ Hz^{-1}\ sr^{-1}} \qquad h\nu < 40\mathrm{keV}. \qquad (2)$$

Unless stated otherwise, in our calculations we use the values $J_{21} = 0.1$ and $\alpha = 0.7$. This normalization of $J_{21}$ is preferred by recent simulations of the Ly$\alpha$ forest (Hernquist et al. 1995; Miralda-Escudé et al. 1995) and is also compatible with the expected emission from the known QSO population (Meiksin & Madau 1993). The power-law index $\alpha = 0.7$ is in accord with quasar spectra, and is also consistent with the fit to the X-ray background spectrum around up to 40 keV (Fabian & Barcons 1992).

The incident spectrum in equation (2) characterizes the background flux before it enters the gas cloud. However, if the gas cloud is optically thick at the Lyman-limit, the shape of the spectrum will be modified due to H-photoionization as the radiation penetrates into the cloud. To mimic the effect of radiative transfer, we modify the spectrum to the form,

$$J_\nu = J_{21} \left(\frac{\nu}{\nu_{\rm H}}\right)^{-\alpha} \exp\{-N_{\rm H}\sigma_{\rm H}(\nu)\} \times 10^{-21} \ \ {\rm erg \ s^{-1} \ cm^{-2} \ Hz^{-1} \ sr^{-1}}, \qquad (3)$$

where $N_{\rm H}$ is the column density (in units cm$^{-2}$) of neutral hydrogen through which the incident flux is processed. For simplicity, we assume that the incident radiation is effectively processed through a uniform slab of gas of length $\ell$ and define

$$N_{\rm H} = f_{\rm H} \left(\frac{\rho}{\mu m_p}\right) \ell = 9.05 \times 10^{20} {\rm cm}^{-2} \times f_{\rm H}^{\frac{1}{2}} \left(\frac{T}{10^3 {\rm K}}\right)^{\frac{1}{2}} \left(\frac{n}{1 {\rm cm}^{-3}}\right)^{\frac{1}{2}} \mu^{-1} \times \frac{\ell}{\ell_{\rm Jeans}} \quad . \qquad (4)$$

Here $f_{\rm H}$ is the neutral hydrogen fraction by number, $\mu$ is the mean molecular weight, $n$ is the total proton number density, and $\ell_{\rm Jeans}$ is the Jeans length for the assumed density, temperature and composition of the slab.

In addressing the question whether the gas cools or heats due to the radiation background at different densities or temperatures, we assume chemical equilibrium. This assumption makes our conclusions independent of the details of hydrodynamic and chemical history of the gas. (We examine below, as a separate question, the time-scale required for establishing the assumed chemical equilibrium). The chemical equilibrium abundances of the various species are solved by integrating equations (1) until the right-hand-side vanishes. Using these abundances, we then calculate the total cooling and heating functions, $\Lambda_{\rm cool}$ and $\Lambda_{\rm heat}$; these quantities are both positive and are expressed in units erg s$^{-1}$ cm$^{-3}$. The interaction between the radiation background and the gas includes the following processes: photoionization and dissociation, Compton scattering, collisional ionization and excitation (including H$_2$), recombination, and Bremsstrahlung emission. Numerical expressions for these processes are listed in Appendix A of Haiman et al. 1995. In addition, for the present calculations we included Compton scattering on the power-law spectrum (cf. eq. 24.42 in Peebles 1993). Compton cooling by the cosmic background radiation is negligible at $z \lesssim 5$.

## 3. Results and Discussion

Under chemical equilibrium and the fixed flux parameters ($J_{21} = 0.1, \alpha = 0.7$), the three parameters that define the state of the gas are its density $\rho$ and temperature $T$, and the column density of the shielding slab, $N_{\rm H}$, in equation (3). At each density, temperature, and column density, the slab width $\ell$ can be expressed in units of the Jeans length using equation (4). We first focus on the sign of $\Lambda_{\rm tot} \equiv \Lambda_{\rm heat} - \Lambda_{\rm cool}$ as a function of these parameters. For reference, the equilibrium $H_2$-fractions as a function of density, temperature, and column density $N_{\rm H}$ are shown in Table 1. Figure 1 shows contours of $\Lambda_{\rm tot} = 0$ in the density-temperature plane for various values of neutral-H column densities ($N_{\rm H}$). At high temperatures ($T \gtrsim 10^4$ K), $H_2$ has no effect, and cooling always dominates through H-excitation. The prominent feature introduced into this figure by $H_2$ is the island of cooling below $T = 10^4$. Within this island, $\Lambda_{\rm heat}$ is dominated by H-photoionization heating, and $\Lambda_{\rm cool}$ is dominated by $H_2$-cooling; the net $\Lambda_{\rm tot}$ is negative. Note that the island diminishes for a thin shielding slab, while its boundaries expand as the column density of the slab grows.

The reasons for the inverse relation between the area of the $H_2$-cooling island and the total transmitted flux $F = \int J_\nu d\nu$ are twofold. The equilibrium $H_2$-fraction $f_{H_2,{\rm eq}}$ is set by a balance between the two-step collisional formation process and the direct photodissociation rate. Hence, $f_{H_2,{\rm eq}}$ increases with decreasing flux, and so does the $H_2$-cooling. In addition, photoionization heating decreases with decreasing flux. Roughly, the scalings are $\Lambda_{\rm cool} \sim 1/\sqrt{F}$ and $\Lambda_{\rm heat} \propto F$, so that $\Lambda_{\rm cool}/\Lambda_{\rm heat} \sim F^{-3/2}$.

The exponential damping factor in equation 3 reduces the total flux, and causes the boundaries of the cooling-island in Figure 1 to expand (note that a similar effect occurs if the overall normalization, $J_{21}$, is lowered). However, it should be borne in mind that as the ionizing flux is reduced, the formation rate of $H_2$ is lowered ($\propto \sqrt{F}$), and chemical equilibrium is less likely to be achieved within a Hubble time. In order to address this issue, we integrated equation (1) as a function of time and calculated the $H_2$ fraction after a Hubble time at $z = 5$, $f_{H_2,{\rm Hub}}$. Initially, the gas was assumed to be neutral and with no $H_2$, i.e. $n_{\rm H^-} = n_{\rm H^+} = n_{\rm He^+} = n_{\rm He^{++}} = n_{H_2} = n_{H_2^+} = n_{e^-} = 0$ and $n_{\rm He}/n_{\rm H} = 0.078$.

Figure 2 shows $f_{H_2,{\rm Hub}}$ and $t_{\rm cool}/t_{\rm dyn}$ as functions of the shielding slab width in units of the Jeans length, for a fixed cloud density and temperature of $n = 1$ cm$^{-3}$ and $T = 10^3$ K. The upper panel demonstrates that as the optical depth of the shielding slab increases, the $H_2$-fraction initially grows while maintaining chemical equilibrium. The value of $f_{H_2,{\rm Hub}}$ peaks at a slab thickness corresponding to $t_{\rm chem} = t_{\rm Hub}$, and declines at yet thicker slabs where it is out of equilibrium. The lower panel of Figure 2 illustrates that the effectiveness of $H_2$-cooling is maintained at much greater optical depths than allowed by the requirement of chemical equilibrium. Indeed, the cooling time-scale remains much shorter than the dynamical time even as the slab thickness exceeds the Jeans length by two orders of magnitude. Even if $H_2$ is out of equilibrium, the increase in the optical depth can increase the ratio $\Lambda_{\rm cool}/\Lambda_{\rm heat}$ sufficiently to turn heating into cooling that is capable of affecting the gas dynamics. For comparison, Figure 2 shows how heating turns into cooling for three different normalizations, $J_{21} = 0.05, 0.10$, and $0.15$.

In conclusion, Figures 1 and 2 demonstrate that as the radiation field penetrates deeper into a gas cloud, it is more likely to cool the gas rather of heat it. There are, however, several limitations on the physical applicability of these results. In the rest of this section we discuss each of these limitations and demonstrate that they play only a minor role in most of the parameter space of the $H_2$-cooling island.

### 3.1. Cooling time vs. dynamical time

The contour-diagram shown in Figure 1 provides only the sign of $\Lambda_{\text{heat}} - \Lambda_{\text{cool}}$. However, the fact that this quantity is negative within the $H_2$-cooling island does not in itself guarantee dynamical consequences. In order for cooling to effect the dynamics of a gas cloud, the cooling time needs to be shorter than the dynamical time. The line $t_{\text{cool}} = t_{\text{dyn}}$ is shown for the five values of column density, $N_H = 0, 10^{18,19,20,21}$ on the density-temperature plane in Figure 1, where we adopt the following definitions for the cooling time,

$$t_{\text{cool}} = \frac{(3/2)nk_B T}{|\Lambda_{\text{heat}} - \Lambda_{\text{cool}}|} = \frac{(3/2)\rho k_B T/\mu m_p}{|\Lambda_{\text{heat}} - \Lambda_{\text{cool}}|} \tag{5}$$

and the dynamical time

$$t_{\text{dyn}} = \frac{1}{\sqrt{6\pi G\rho}}. \tag{6}$$

It is apparent from Figure 2 that most of the $H_2$-cooling island, except for its high-density high-temperature corner, lies in the region where $t_{\text{cool}} < t_{\text{dyn}}$. This implies that $H_2$-cooling in most cases is indeed efficient enough to affect the dynamics of the gas.

### 3.2. Chemical equilibrium vs. Hubble time

Figure 1 assumes chemical equilibrium. If the time-scale to reach chemical equilibrium, $t_{\text{chem}}$, is too long, the $H_2$-fraction could be significantly lower than its equilibrium value $f_{H_2,\text{eq}}$. Although the chemical reaction-network in equation (1) does not possess a single time-constant that characterizes relaxation to chemical equilibrium, the time-scale for $H_2$ to reach equilibrium is roughly proportional to the square of the number density of free electrons, i.e. $\propto 1/F$.

We compare $t_{\text{chem}}$ and the Hubble time at $z = 5$, $t_H = 2.0 \times 10^{16}$ s (assuming $h = 0.5$ and $\Omega = 1$). Although the chemical time-scale could also be compared to the dynamical time, this would be less relevant for virialized objects that do not undergo bulk motions in a dynamical time. Figure 1 shows (dashed) lines of $t_{\rm chem} = t_{\rm Hub}$ for three values of the column density, $N_{\rm H} = 10^{19,20,21}$. For $N_{\rm H} = 0$ and $10^{18}$, $H_2$ is in equilibrium everywhere in Figure 1. In the regions where $t_{\rm chem} < t_{\rm Hub}$, chemical equilibrium can, in principle, be established. In the regions where $t_{\rm chem} > t_{\rm H}$, the $H_2$-fraction reached within a Hubble time $f_{\rm H_2,Hub}$ depends on the initial abundances and is always lower than the equilibrium abundance $f_{\rm H_2,eq}$ if initially $f_{\rm H_2}=0$. Nevertheless, we find that the contour-diagram is practically *unchanged* if $f_{\rm H_2,Hub}$ is used in place of $f_{\rm H_2,eq}$ to calculate $\Lambda_{\rm tot}$. The out-of-equilibrium $H_2$ fraction becomes dynamically important, with $\Lambda_{\rm tot} < 0$ and $t_{\rm cool}/t_{\rm dyn} < 1$ within a Hubble time, even for the lowest column density shown in Figure 1, $N_{\rm H} = 10^{21}$ cm$^{-2}$.

### 3.3. Chemical equilibrium vs. thermal equilibrium

In realistic circumstances the approach to thermal and chemical equilibrium may be coupled. This coupling could affect the interpretation of Figures 1 and 2. Here, we will consider the regime where both the thermal and the chemical time-scales are shorter than the dynamical time. We therefore keep the density fixed, and consider the case $n = 100 {\rm cm}^{-3}$.

If the initial $H_2$ abundance is small, then as soon as the background radiation turns on the gas will heat. The subsequent thermal evolution of the gas, however, depends on the competition between the photo-ionization heating rate and the increase in the $H_2$ cooling rate. Both the initial heating rate and the initial $H_2$-formation rate increase with increasing incident flux. However, the heating rate is proportional to the total flux $F$, while the $H_2$-formation rate grows only as $\sqrt{F}$. The $H_2$-formation rate is also sensitive to the gas temperature; it increases with temperature up to $\sim 3,000$K but then decreases abruptly at yet higher temperatures because of collisional dissociation. Thus, the relevant question is whether photoionization heating raises the gas temperature above $\sim 3,000$K before $H_2$ is formed in sufficient quantities to reverse the sign of $\Lambda_{\rm heat} - \Lambda_{\rm cool}$. As we show below, the answer to this question is positive for high incident fluxes and negative for low fluxes.

We solved the coupled time-evolution of the temperature and the relative abundances during the approach to thermal and chemical equilibrium. This was done by integrating equation (1), starting from $T = 10^3$ K and allowing the temperature and the rate coefficients $k_{\rm lm}$ to evolve at each timestep according to $\Lambda_{\rm heat} - \Lambda_{\rm cool}$. The initial abundances were chosen as in Figure 2. Figure 3 illustrates the sensitivity of the resulting temperature and $H_2$-fraction to the incident flux, characterized by the slab column density $N_{\rm H}$. For a higher flux ($N_{\rm H} = 0$, upper panel), $\Lambda_{\rm heat}$ remains roughly constant since the ionization fraction is low but $\Lambda_{\rm cool}$ is increasing as more $H_2$ is produced. When the temperature reaches $\sim 3,000$ K, molecular hydrogen is destroyed. The final equilibrium H$_2$-fraction is correspondingly low, $f_{H_2,eq} = 4.0 \times 10^{-6}$, as the final equilibrium temperature reaches the high value of 8,440 K. The lower panel shows a case when the incident flux is lower, due to a thick slab $N_H = 10^{21}$. Here the H$_2$-fraction increases sufficiently rapidly so as not to allow the temperature to reach $\sim 3,000$ K. In this case, the final equilibrium H$_2$-fraction is high, $f_{H_2,eq} = 2.5 \times 10^{-2}$, and the equilibrium temperature is low, 290 K. In fact, the equilibrium temperature in this case is *below* the initial temperature of 1,000 K, implying that a net cooling feedback due to molecular hydrogen is possible.

### 3.4. Additional Processes

In addition to the time-scale issues addressed in § 3.1-3.3, we examined three processes that could invalidate our conclusions. We found all three processes to be insignificant, and therefore only comment on them briefly below.

3.5.1 H$_2$ *destruction by the Solomon process*

H$_2$-molecules can be dissociated by photons with energies below 13.6eV by a two-step process, as first suggested by Solomon in 1965 (cf. Field et al. 1966) and later shown quantitatively by Stecher & Williams (1967). The incident number flux of photons is likely to be higher below the H-ionization threshold of 13.6eV than above it, especially if the radiation goes through a slab of neutral hydrogen which absorbs photons only above 13.6eV (cf. eq. [3]). Nevertheless, we find that because of its small cross-section $\sigma_{sw} \sim 3 \times (10^{-21}-10^{-22})$ cm$^2$, the Solomon process does not effect the contour maps shown in Figure 1. As long as there is a high-energy component to the incident spectrum, the Solomon process is unable to compete with the direct photodissociation of H$_2$ by the high-energy photons.

3.5.2 *Fast electrons*

H$_2$-molecules could also be dissociated by fast electrons that arise as the product of ionization by energetic photons ($h\nu \sim$ 1keV). In fact, depending on the ionization fraction (which determines the drag on the fast electrons) one fast electron can dissociate several H$_2$-molecules. Xu and McCray (1991) give the fractional energy deposition into thermalization, excitations, and dissociations, by fast electrons for various energies and ionization fractions. Based on these fractions, we conclude that even though as much as 30% of the electron energy can go into dissociations, the number of H$_2$-destructions by fast electrons is still small ($< 15\%$) in comparison with the number of direct H$_2$-photoionizations. Consequently, the boundaries in the contour map (Fig. 1) would not change even if multiple-dissociations by fast electrons were included.

*3.5.3 Self-absorption of infrared photons*

The $H_2$-cooling could, in principle, be suppressed if the gas cloud were optically-thick to its emitted infrared-line photons. $H_2$-cooling occurs mainly due to radiative de-excitation from excited rotational and vibrational levels of the electronic ground state. We obtained cross-sections for these transitions from the transition probabilities given in Turner et al. 1977, and found them to be smaller than $\sim 10^{-26}$ cm$^2$. Given these small cross-sections and the typically small $H_2$-fraction ($\lesssim 0.01$), we conclude that an exceedingly high H-column density, $\sim 10^{28}$ cm$^{-2}$, is necessary to absorb the IR cooling photons.

## 4. Cosmological Implications

In summary, we find that a UV irradiation of a cold gas cloud with a density $\gtrsim 1$ cm$^{-3}$ can trigger excess $H_2$-cooling on a time-scale shorter than its dynamical time. The effect is more pronounced in the inner cores of massive gas clouds which are shielded from photoionization heating by the surrounding material (cf. Fig. 1). Contrary to naive expectations, the existence of a strong ionizing background at redshifts $z \sim 2-4$, could have *accelerated* the collapse and fragmentation of dense objects, rather than increased their thermal pressure support. This $H_2$-cooling could have lowered the Jeans-mass threshold for fragmentation and led to enhanced star-formation rates in these objects. Since the UV background is found to decline to low values at low-redshifts (Kulkarni & Fall 1993), this enhancement in the fragmentation efficiency may have been a unique feature of the universe at redshifts $z \sim 2-4$.

The $H_2$-cooling effect may also have important implications on the collapse of objects, and the duration of the reionization epoch at higher redshifts. The cooling island in Figure 1 extends down to a density corresponding to a redshift $(1+z) \gtrsim 30 \left(\Omega_b h^2/0.025\right)^{-1/3} (1+\Delta/200)^{-1/3}$, where $\Delta$ is the overdensity in a collapsing region and $\Omega_b$ is the baryonic density parameter. In particular, if the very first objects to "light up" triggered the cooling effect discussed here, they could have accelerated the formation of additional ionizing sources, and reionization would have been completed abruptly. In contrast, if the first sources did not exert this positive feedback (e.g. because their UV spectrum was stellar, and did not extend towards the X-ray band to trigger the formation of $H_2$), then the photoionization of the IGM must have taken longer, with the Jeans mass rising continuously up to $10^{8-9} M_\odot$. These two scenarios differ in the amount of small-scale structure formed by the end of the reionization epoch.

## Acknowledgements

We thank Alex Dalgarno and Min Yan for helpful discussions of the chemistry involved. ZH acknowledges financial support from a Cambridge European Trust ORS award and the Isaac Newton Studentship in Cambridge. AL acknowledges support from the NASA ATP grant NAG5-3085.

Table 1: The equilibrium $H_2$-fraction $f_{H_2,eq} = 2n_{H_2}/(n_H + n_{H^-} + n_{H^+} + 2n_{H_2} + 2n_{H_2^+})$ as a function of total number density $n$, temperature $T$, and column density $N_H$. For each column density, the five rows correspond to five different temperatures $2.0 \leq \log(T/K) \leq 4.0$, the eight columns to the number densities $-1.0 \leq \log(n/\text{cm}^{-3}) \leq 6.0$.

| $n_H = 0$ | -1.0 | 0.0 | 1.0 | 2.0 | 3.0 | 4.0 | 5.0 | 6.0 |
|---|---|---|---|---|---|---|---|---|
| 2.0 | 0.48E-7 | 0.11E-5 | 0.15E-4 | 0.17E-3 | 0.18E-2 | 0.19E-1 | 0.14E+0 | 0.50E+0 |
| 2.5 | 0.89E-7 | 0.28E-5 | 0.41E-4 | 0.46E-3 | 0.48E-2 | 0.45E-1 | 0.27E+0 | 0.64E+0 |
| 3.0 | 0.13E-6 | 0.67E-5 | 0.11E-3 | 0.13E-2 | 0.13E-1 | 0.10E+0 | 0.42E+0 | 0.76E+0 |
| 3.5 | 0.12E-6 | 0.59E-5 | 0.47E-4 | 0.19E-3 | 0.60E-3 | 0.17E-2 | 0.50E-3 | 0.93E-4 |
| 4.0 | 0.29E-8 | 0.82E-7 | 0.66E-6 | 0.30E-5 | 0.11E-4 | 0.21E-4 | 0.26E-6 | 0.48E-7 |
| $n_H = 10^{18}$ | -1.0 | 0.0 | 1.0 | 2.0 | 3.0 | 4.0 | 5.0 | 6.0 |
| 2.0 | 0.69E-6 | 0.10E-4 | 0.12E-3 | 0.12E-2 | 0.11E-1 | 0.92E-1 | 0.40E+0 | 0.74E+0 |
| 2.5 | 0.17E-5 | 0.28E-4 | 0.31E-3 | 0.30E-2 | 0.27E-1 | 0.19E+0 | 0.55E+0 | 0.83E+0 |
| 3.0 | 0.38E-5 | 0.74E-4 | 0.83E-3 | 0.77E-2 | 0.64E-1 | 0.33E+0 | 0.69E+0 | 0.89E+0 |
| 3.5 | 0.43E-5 | 0.47E-4 | 0.21E-3 | 0.69E-3 | 0.21E-2 | 0.34E-2 | 0.21E-3 | 0.36E-4 |
| 4.0 | 0.75E-7 | 0.64E-6 | 0.30E-5 | 0.11E-4 | 0.37E-4 | 0.17E-4 | 0.14E-6 | 0.36E-7 |
| $n_H = 10^{19}$ | -1.0 | 0.0 | 1.0 | 2.0 | 3.0 | 4.0 | 5.0 | 6.0 |
| 2.0 | 0.26E-4 | 0.30E-3 | 0.26E-2 | 0.23E-1 | 0.16E+0 | 0.51E+0 | 0.80E+0 | 0.93E+0 |
| 2.5 | 0.64E-4 | 0.77E-3 | 0.67E-2 | 0.54E-1 | 0.29E+0 | 0.66E+0 | 0.87E+0 | 0.96E+0 |
| 3.0 | 0.16E-3 | 0.20E-2 | 0.17E-1 | 0.12E+0 | 0.45E+0 | 0.77E+0 | 0.92E+0 | 0.97E+0 |
| 3.5 | 0.76E-4 | 0.39E-3 | 0.13E-2 | 0.39E-2 | 0.89E-2 | 0.20E-2 | 0.66E-4 | 0.11E-4 |
| 4.0 | 0.98E-6 | 0.45E-5 | 0.17E-4 | 0.55E-4 | 0.11E-3 | 0.91E-5 | 0.95E-7 | 0.34E-7 |
| $n_H = 10^{20}$ | -1.0 | 0.0 | 1.0 | 2.0 | 3.0 | 4.0 | 5.0 | 6.0 |
| 2.0 | 0.47E-3 | 0.48E-2 | 0.34E-1 | 0.17E+0 | 0.46E+0 | 0.75E+0 | 0.91E+0 | 0.97E+0 |
| 2.5 | 0.12E-2 | 0.12E-1 | 0.80E-1 | 0.32E+0 | 0.65E+0 | 0.87E+0 | 0.95E+0 | 0.99E+0 |
| 3.0 | 0.29E-2 | 0.29E-1 | 0.17E+0 | 0.50E+0 | 0.79E+0 | 0.93E+0 | 0.98E+0 | 0.99E+0 |
| 3.5 | 0.49E-3 | 0.22E-2 | 0.70E-2 | 0.15E-1 | 0.83E-2 | 0.69E-3 | 0.21E-4 | 0.36E-5 |
| 4.0 | 0.59E-5 | 0.26E-4 | 0.79E-4 | 0.15E-3 | 0.13E-3 | 0.67E-5 | 0.90E-7 | 0.33E-7 |
| $n_H = 10^{21}$ | -1.0 | 0.0 | 1.0 | 2.0 | 3.0 | 4.0 | 5.0 | 6.0 |
| 2.0 | 0.21E-2 | 0.12E-1 | 0.42E-1 | 0.13E+0 | 0.32E+0 | 0.59E+0 | 0.82E+0 | 0.94E+0 |
| 2.5 | 0.52E-2 | 0.30E-1 | 0.10E+0 | 0.27E+0 | 0.54E+0 | 0.79E+0 | 0.92E+0 | 0.97E+0 |
| 3.0 | 0.13E-1 | 0.71E-1 | 0.22E+0 | 0.48E+0 | 0.75E+0 | 0.91E+0 | 0.97E+0 | 0.99E+0 |
| 3.5 | 0.26E-2 | 0.12E-1 | 0.22E-1 | 0.12E-1 | 0.33E-2 | 0.24E-3 | 0.75E-5 | 0.13E-5 |
| 4.0 | 0.30E-4 | 0.10E-3 | 0.18E-3 | 0.18E-3 | 0.12E-3 | 0.64E-5 | 0.89E-7 | 0.33E-7 |

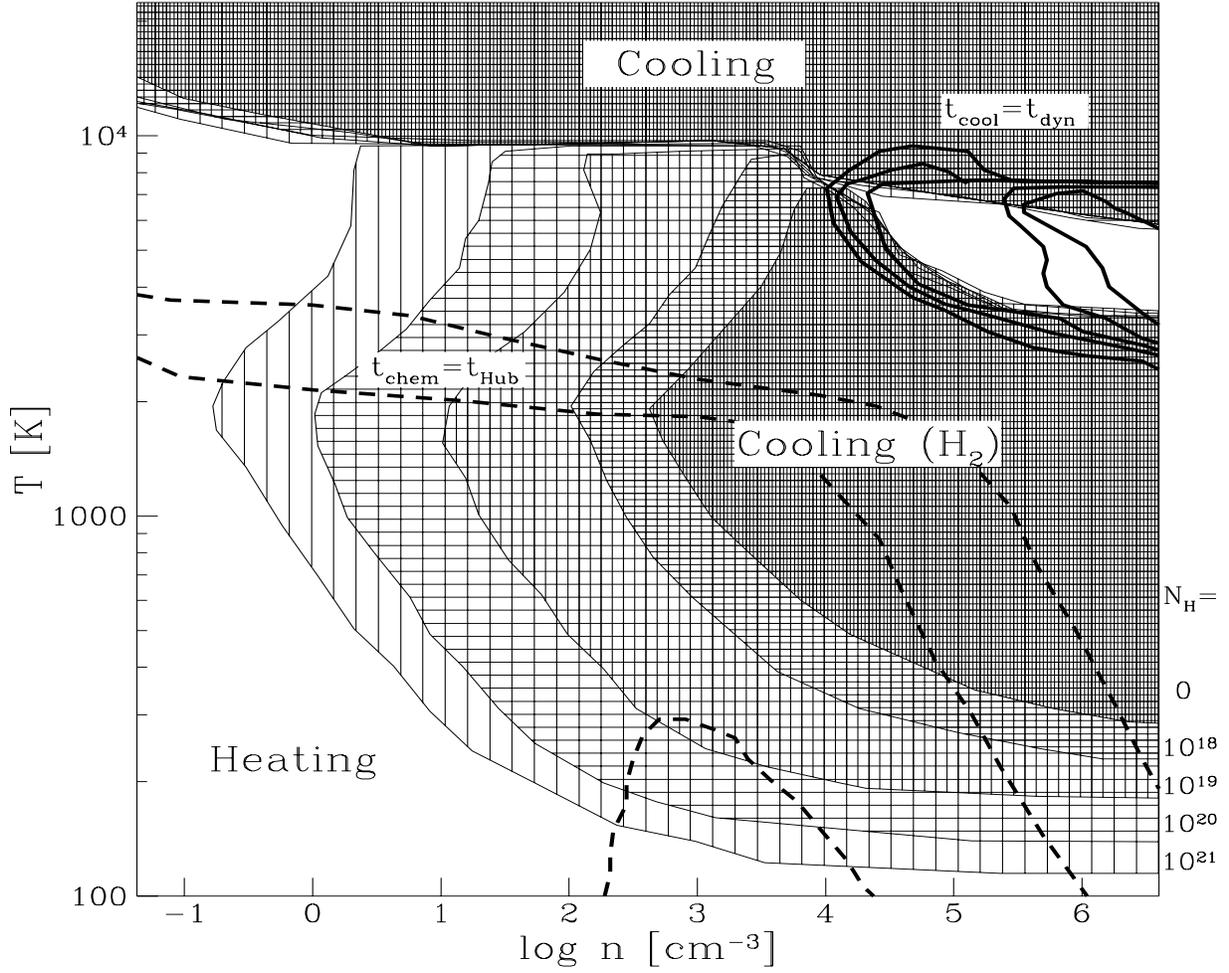

Fig. 1.— Zero energy exchange contours ($\Lambda_{\rm heat} - \Lambda_{\rm cool} = 0$) for five different column-densities $N_{\rm H} = 0, 10^{18,19,20,21}$ cm$^{-2}$. The five solid lines in the upper right corner show corresponding lines of $t_{\rm cool} = t_{\rm dyn}$ (the right-most line corresponds to $N_{\rm H} = 0$, the left-most line to $N_{\rm H} = 10^{21}$) . Going to the left, the area enclosed by the lines defines increasingly larger regions of ineffective cooling ($t_{\rm cool} > t_{\rm dyn}$). The three dashed curves show lines of $t_{\rm chem} = t_{\rm Hub}$ for $N_{\rm H} = 10^{19}$ (lower curve) and $N_{\rm H} = 10^{20,21}$ (upper two curves). Above these lines, chemical equilibrium is established in a Hubble time. For $N_{\rm H} = 0$ and $10^{18}$, the lines are off the plot.

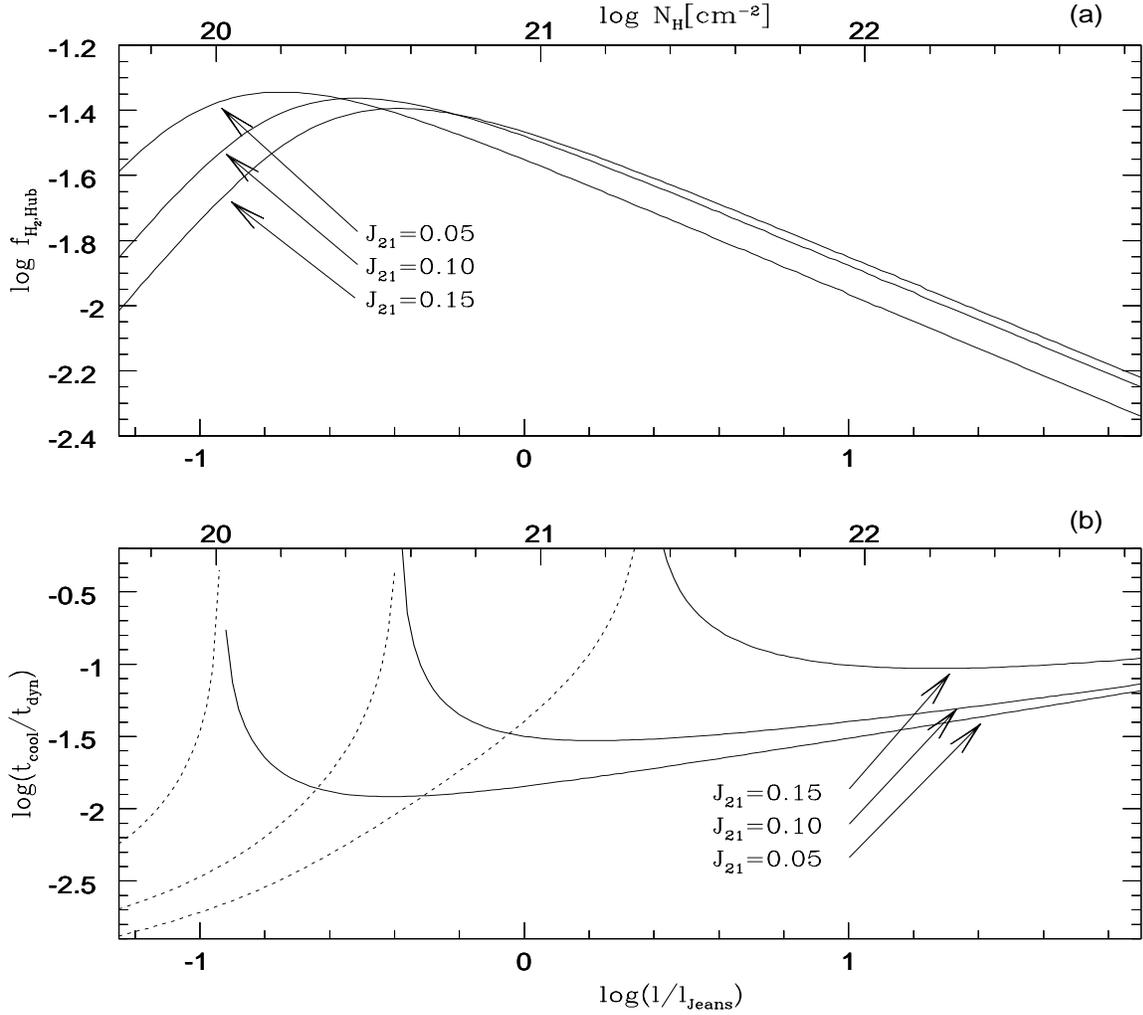

Fig. 2.— Effects of shielding by a slab of thickness $\ell$ on the $H_2$-fraction reached in a Hubble time (upper panel), and on the ratio between the cooling time and the dynamical time (lower panel). In the lower panel, cooling is denoted by a solid line and heating by a dotted line. The gas is assumed to have a density $1\ \text{cm}^{-3}$ and a temperature $T = 10^3$ K. The horizontal axis describes the thickness of the shielding slab in units of the Jeans length (bottom label) and in the corresponding column density $\log N_H$ (top label).

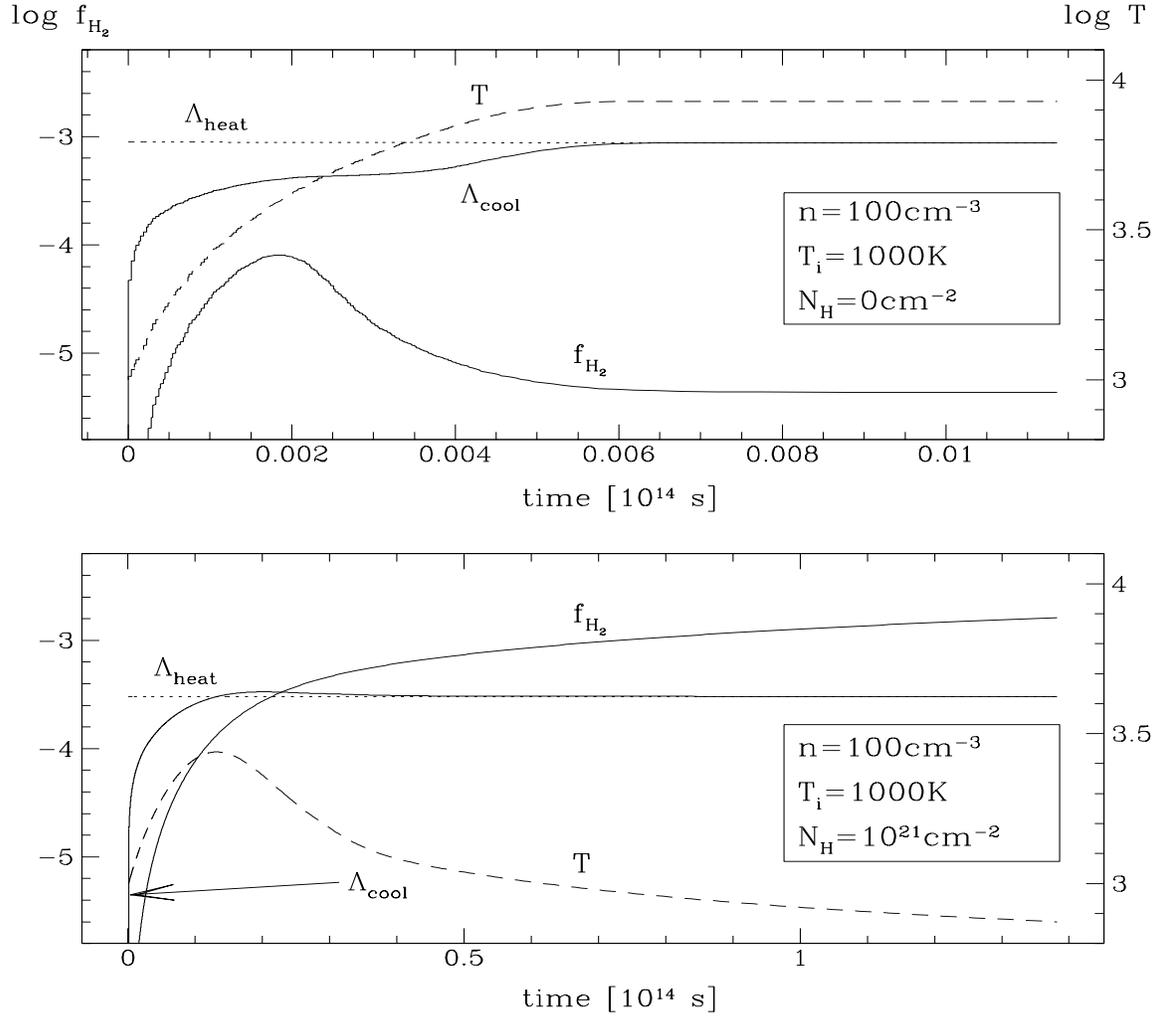

Fig. 3.— The approach to equilibrium when thermal and chemical evolution are coupled. The vertical axis shows log of the $H_2$-fraction on the left, and log of the temperature in K on the right. The heating and cooling functions are arbitrarily normalized.